\documentclass[conference]{IEEEtran}
\usepackage{booktabs} 

\usepackage{graphics}
\usepackage{epstopdf}
\usepackage{psfrag}
\usepackage{textcomp}
\usepackage{listings}
\usepackage{float}
\usepackage{epsfig}
\usepackage{amsmath}
\usepackage{amssymb}
\usepackage{amsthm}
\usepackage[linesnumbered,ruled]{algorithm2e}
\usepackage{algorithmic}
\usepackage[hang,flushmargin]{footmisc}
\usepackage{url}

\usepackage{tikz}
\usetikzlibrary{positioning}
\usepackage{hhline}
\usepackage{multirow}
\usepackage{color}

\usepackage{xcolor}
\usepackage{listings}

\definecolor{mGreen}{rgb}{0,0.6,0}
\definecolor{mRed}{rgb}{0.6,0,0}
\definecolor{mBlue}{rgb}{0,0,0.6}
\definecolor{mGray}{rgb}{0.5,0.5,0.5}
\definecolor{mPurple}{rgb}{0.58,0,0.82}
\definecolor{backgroundColour}{rgb}{0.92,0.92,0.92}
\definecolor{light-gray}{gray}{0.95}
\definecolor{ballblue}{rgb}{0.13, 0.67, 0.8}
\definecolor{burntorange}{rgb}{0.8, 0.33, 0.0}

\lstdefinestyle{CStyle}{
    backgroundcolor=\color{light-gray},   
    commentstyle=\color{burntorange},
    numberstyle=\tiny,
    stringstyle=\color{mGreen},
    basicstyle=\ttfamily\footnotesize,
    breakatwhitespace=false,
    xleftmargin=.1in,
    breaklines=true,                 
    captionpos=b,                    
    keepspaces=true,                 
    numbers=left,                    
    numbersep=5pt,                  
    showspaces=false,                
    showstringspaces=false,
    showtabs=false,                  
    tabsize=2,
    keywordstyle=\color{mPurple},
    deletekeywords={pragma,for,int,while,double},
    morekeywords={pragma,omp,master,parallel,num\_threads},
    keywordstyle=[2]\color{mBlue},
    keywords=[2]{var,int,bool,atomic,sync,const,class,domain,proc,double}, 
  	keywordstyle=[3]\color{mRed},
    keywords=[3]{for,in,forall,coforall,if,while},
}

\newcommand{\todo}[1]{}
\renewcommand{\todo}[1]{{\color{red} TODO: {#1}}}

\usepackage[tight,footnotesize]{subfigure}

\usepackage{float}
\begin{document}


%
%

\title{An Empirical Evaluation of Allgatherv on Multi-GPU Systems}


\author{\IEEEauthorblockN{Thomas B. Rolinger\IEEEauthorrefmark{1}\IEEEauthorrefmark{2}, Tyler A. Simon\IEEEauthorrefmark{1}, and Christopher D. Krieger\IEEEauthorrefmark{1}}
\IEEEauthorblockA{\IEEEauthorrefmark{1}
Laboratory for Physical Sciences, University of Maryland, College Park, MD USA
\IEEEauthorblockA{\IEEEauthorrefmark{2}
Department of Computer Science, University of Maryland, College Park, MD USA}
tbrolin@cs.umd.edu, \{tasimon,krieger\}@lps.umd.edu}}

\maketitle
\begin{abstract}
Applications for deep learning and big data analytics have compute and memory requirements that exceed the limits of a single GPU. 
However, effectively scaling out an application to multiple GPUs is challenging due to the complexities of communication between the GPUs, particularly for collective communication with irregular message sizes.
In this work, we provide a performance evaluation of the Allgatherv routine on multi-GPU systems, focusing on GPU network topology and the communication library used.
We present results from the OSU-micro benchmark as well as conduct a case study for sparse tensor factorization, one application that uses Allgatherv with highly irregular message sizes.
We extend our existing tensor factorization tool to run on systems with different node counts and varying number of GPUs per node. 
We then evaluate the communication performance of our tool when using traditional MPI, CUDA-aware MVAPICH and NCCL across a suite of real-world data sets on three different systems: a 16-node cluster with one GPU per node, NVIDIA's DGX-1 with 8 GPUs and Cray's CS-Storm with 16 GPUs.
Our results show that irregularity in the tensor data sets produce trends that contradict those in the OSU micro-benchmark, as well as trends that are absent from the benchmark.
\end{abstract}


\section{Introduction}
\label{sec:intro}

In high performance computing, an application's memory requirements often exceed the limits of a single compute node, driving the development of distributed-memory algorithms and implementations.
Large-memory applications that leverage graphics processing units (GPUs) face greater challenges because the amount of available memory on a GPU is significantly less than that of a compute node.
In these cases, it is imperative to effectively use multiple GPUs together, whether distributed across multiple nodes, on the same node, or both~\cite{CNN_MultiGPU,SpMV_MultiGPU,Lasso_MultiGPU}.

Scaling an application out to multiple GPUs can be a difficult task due to the complexities of communication between the GPUs~\cite{evalInterNodeGPU}.
As communication costs can be the performance bottleneck for many applications~\cite{mpichOptimizations}, achieving good communication performance is imperative.
The key factors in multi-GPU communication performance are:

\begin{itemize}
\item GPU network topology of the system
\item Communication pattern of the application
\item Size of messages
\item Underlying communication library
\end{itemize}
Understanding how these features impact performance can significantly aid in the development of efficient multi-GPU applications and systems.

In this paper, we focus on collective communication between GPUs with irregular message sizes, specifically the \texttt{Allgatherv} routine.
The \texttt{Allgatherv} routine is present in many applications, including matrix multiplication, LU decomposition, 2D parallel breadth first search, and conjugate gradient solvers~\cite{nonUniformMPI,2dBFS}.
To the best of our knowledge, current benchmarks for GPU communication performance, such as the Ohio State University (OSU) micro-benchmarks\footnote{http://mvapich.cse.ohio-state.edu/benchmarks/}, do not evaluate \texttt{Allgatherv} with irregular message sizes and instead use fixed size messages.

It is difficult to implement meaningful benchmarks for irregular behavior that generalize to a wide range of applications. To this end, we conduct a case study of a specific irregular application, namely \textit{sparse tensor factorization}, which uses \texttt{Allgatherv} for communication.
Tensor factorization is emerging as a popular technique in big-data analytics as it provides a natural way to model and extract patterns from large, sparse multi-way data.
Furthermore, the irregular collective communication characteristics we observe in tensor factorization can also be found in other applications that involve sparse linear algebra.
Therefore, our findings can be applied to applications beyond tensor factorization.

In this paper, we provide insight into the following topics:

\begin{itemize}
\item How much performance do we gain from using a specialized single node, multi-GPU system over a traditional cluster with respect to \texttt{Allgatherv} performance?
\item Can special-purpose GPU communication libraries geared towards dense multi-GPU systems and regular workloads be competitive with more general-purpose libraries on traditional clusters?
\item How does message size irregularity factor into the communication performance with respect to a system's GPU network topology and the communication library used?
\end{itemize}

In addressing these questions, we provide the following contributions:
\begin{enumerate}
\item An evaluation of \texttt{Allgatherv} communication costs for a variety of multi-GPU topologies using current state-of-the art GPU communication libraries. 
The systems we evaluate include a traditional cluster with one GPU per node, NVIDIA's DGX-1 with 8 GPUs connected by NVIDIA's NVLink interconnect, and Cray's CS-Storm with 16 GPUs connected in pairs by NVLink.
For communications libraries, we focus on the MVAPICH MPI library with and without CUDA support enabled as well as the NVIDIA Collective Communications Library (NCCL).
We evaluate both inter- and intra-node communication performance using the OSU benchmark, which we extend to allow for NCCL to be used.
\item A case study for communication runtime in sparse tensor factorization. 
We extend an existing tensor factorization tool to run on multi-GPU systems and compare traditional MPI, CUDA-aware MVAPICH-GDR and NCCL as the communication library.
To the best of our knowledge, this is the first approach to tensor factorization that executes across multiple GPUs and performs communication explicitly between the GPUs.
We perform experiments on real-world data sets that exhibit a high degree of irregularity in message sizes, with as much as a 25,400x difference between the smallest and largest message size within a given data set.
\end{enumerate}

The rest of this paper is organized as follows.
In Section \ref{sec:gpuComm}, we describe two popular approaches for collective communication between GPUs, namely CUDA-aware MPI and NCCL.
We provide a brief overview of tensor factorization in Section \ref{sec:tensors}, as well as a description of the specific implementation that we use in our study.
In Section \ref{sec:related}, we discuss related work for both GPU communication and tensor factorization.
Section \ref{sec:eval} describes the multi-GPU systems we executed on and then presents the results of our experiments.
Finally, we present concluding remarks and future work in Section \ref{sec:concl}.

\section{GPU Collective Communication}
\label{sec:gpuComm}
For multi-GPU systems, optimizing collective communication is challenging  due to the complexities in GPU network topologies.
In multi-GPU systems, a variety of interconnects are used to network multiple GPUs together, each with different bandwidths and latencies.
Because of this, communication between a pair of GPUs may be much faster and more efficient than between another otherwise identical pair of GPUs.
This presents a challenge to the underlying communication library to determine the optimal communication paths between any two GPUs.

In this section, we discuss two popular approaches for performing collective communication both between GPUs within the same node and across multiple nodes: CUDA-aware MPI and NVIDIA's Collective Communications Library (NCCL).

\subsection{CUDA-aware MPI}
\label{sec:cudaMPI}
Combining MPI with CUDA has become a popular strategy for solving problems that consume more memory than is available on a single GPU~\cite{cudaMPIExample,cudaMPIExample2}.
For such problems, the data to be shared among the MPI ranks is often stored on the GPUs.
In the past, MPI libraries had no direct support for CUDA and the task of communicating data between GPUs was left to the user, requiring explicit memory-staging through the host to send data from one GPU to another.
Sending data required multiple data copies, first from the GPU to the CPU, then to a network card, transmission over an interconnect, a copy from the network card to the CPU, and finally from the CPU to the GPU. The result was overall poor performance~\cite{MVAPICH_GPUIB}.

The term \emph{CUDA-aware MPI} refers to MPI implementations that enable more efficient communication between GPUs by leveraging features in GPU hardware. 
The API of a CUDA-aware MPI implementation does not need to be altered to support GPU communication due to the introduction of Unified Virtual Addressing (UVA) in CUDA 4.0.
UVA treats the host memory on a single node and the memory of the GPUs on that node as one virtual address space.
Because of this, determining whether a buffer passed to an MPI call resides on the host or GPU can be done by observing the high-order bits of its address.
Current MPI implementations that are CUDA-aware include MVAPICH, OpenMPI, CRAY MPI and SGI MPI.
In this work, we use MVAPICH as our MPI implementation.
Details of MVAPICH's approach to CUDA-aware MPI can be found in~\cite{MVAPICH_GPUIB,MVAPICH_IPC,MVAPICH-GDR}.

Between multiple GPUs on a single node, CUDA-aware MPI can take advantage of NVIDIA's GPUDirect Peer-to-Peer (P2P)  and CUDA IPC features to copy data between two GPU memories without going through the host.
This is accomplished by utilizing UVA and the available interconnects between GPUs, such as PCIe or NVLink.
GPUDirect RDMA (GDR) is a feature introduced in CUDA 5.0 and allows for efficient communication for GPUs residing on separate hosts, such as the nodes in a cluster.
With GDR, GPUs have direct read and write access to the network adapters on the host, eliminating the need to stage data through the host for inter-node communication.

\subsection{NVIDIA Collective Communications Library}
\label{sec:nccl}
NVIDIA's Collective Communications Library (NCCL)\footnote{https://developer.nvidia.com/nccl} provides multi-GPU collective communication routines that are bandwidth-optimized for NVIDIA GPUs.
The supported collectives include all-gather, all-reduce, broadcast, reduce and reduce-scatter.
Originally, NCCL was limited to single node multi-GPU systems.
However, the latest version of NCCL provides the capability for inter-node communication, allowing for data transfers between GPUs on separated nodes~\cite{NCCL2}.

NCCL's target application is deep learning, where every GPU pushes updates during the training phase to all other GPUs.
In such an application, the communication is almost always collective and the sizes of the buffers being sent are typically large and fixed.
Because of these characteristics, NCCL is currently limited to regular collective communication and focuses on optimizing bandwidth over latency to better service the large message sizes.
CUDA-aware MPI does not target specific GPU network topologies and due to MPI's underlying design, it is capable of scaling out to potentially hundreds or thousands of nodes and GPUs.
On the other hand, NCCL is a special purpose communication library geared towards systems such as NVIDIA's DGX-1, which has 8 GPUs within a single node.

NCCL provides automatic topology detection to determine the optimal communication path between the GPUs in a system.
This provides NCCL with an advantage over CUDA-aware MPI on systems such as the DGX-1 that have complicated topologies.
For example, in the right diagram of Figure \ref{fig:diagrams}, the P100 GPU with ID 0 can communicate with GPUs 5, 6 and 7 by traversing two NVLink connections or by going through the PCIe network and the host.
CUDA-aware MPI will not be able to detect that GPU 0 can reach those GPUs via NVLink because GPUDirect P2P is not supported between those GPUs and will default to using PCIe and the host.
However, NCCL's topology detection does not rely entirely on whether GPUDirect P2P is supported between GPUs and will determine that the optimal path between GPU 0 and the others is via two NVLink hops.
Inter-node communication is accomplished using a combination of GDR, Ethernet sockets and Infiniband verbs.

\section{Tensor Factorization}
\label{sec:tensors}
Some data analytics process large amounts of sparse, high dimensional data, such as that produced in signal processing~\cite{de1998matrix} and geospatial analysis~\cite{hpecGeo}.
Within such areas, analysts are interested in discovering previously unknown relationships between elements in the data.
A \textit{tensor} is a matrix extended to three or more dimensions and is a natural way to model this large, sparse multi-way data.
\textit{Tensor factorization}, often called tensor decomposition, is an extension of two dimensional singular value decomposition (SVD) to tensors~\cite{KoldaSIAM09}.

In this section, we present our existing multi-GPU based tensor factorization tool, ReFacTo, as well as the steps taken to extend ReFacTo for this work to support multi-GPU communication.

\subsection{ReFacTo}
\label{sec:refacto}
The Canonical Decomposition/Parallel Factorization (CP) algorithm is widely used to perform tensor factorization.
CP computes a set of $R$ components that sum to an approximation of the original tensor, where $R$ is referred to as the $rank$ of the decomposition.
While there are several methods for computing these components, we use the most common method, alternating least squares (ALS).
Our code, ReFacTo~\cite{rolingerHPEC2017}, is a GPU-extension of DFacTo~\cite{DFacto}, a CPU-based distributed memory implementation of CP-ALS. ReFacTo employs a CUDA+MPI programming model, running across distributed nodes in a cluster and using a single GPU on each node to perform sparse matrix times vector (SpMV) operations via NVIDIA's cuSPARSE library. 

Like DFacTo, ReFacTo uses a coarse-grained data decomposition to assign contiguous slices of the tensor to each MPI rank, which are then responsible for computing the corresponding rows in a set of dense matrices.
Each mode of the tensor is associated with one of these matrices, where the dimensions of the matrix are determined by the length of the mode and the rank of the decomposition. 
We refer to these matrices as \textit{factor matrices}.
Due to ReFacTo/DFacTo's implementation of CP-ALS, each MPI rank is required to store a full copy of the factor matrices and relies on the other ranks to compute the rows that are assigned to them.
The rows are communicated between the ranks using the MPI \texttt{Allgatherv} routine.
We can formalize the communication volume per MPI rank as $(\frac{M_{i}}{P})\log_{2}P$, where $M_{i}$ is the number of rows in the $i^{th}$ mode's factor matrix assigned to a MPI rank and $P$ is the total number of MPI ranks.

The size of each mode in a given tensor is proportional to the
total amount of data that is communicated within ReFacTo.
The sizes of the messages sent throughout a decomposition can vary by orders of magnitude due to the size disparity of the tensor's modes.
Within each mode of the tensor, the size of the messages exchanged within a single \texttt{Allgatherv} call can also vary widely due to the number of rows assigned to each MPI rank, which is influenced by the distribution of non-zeros in the original tensor.
Because of these factors, the messages sent in ReFacTo are often very different in size, ranging from less than 1MB to more than 450MB (see Table \ref{tab:datasets} for explicit examples).

\subsection{Extending ReFacTo for Multi-GPU Communication}
\label{sec:refactEXT}
For this work, we modify ReFacTo by addressing some of the performance challenges identified in prior work~\cite{rolingerHPEC2017} as well as extend ReFacTo to run on a variety of multi-GPU systems and perform GPU-to-GPU communication.
The fact that the factor matrices need to be stored in device memory for GPU communication but also be accessed on the host for other CP-ALS routines presents a challenge for effectively utilizing GPU communication libraries.
Therefore, we ported all of CP-ALS onto the GPU, rather than only the SpMV operations.
While some routines in CP-ALS contribute very little to the overall CP-ALS runtime and may not port well onto the GPU, it is a net performance gain to execute these routines on the GPU instead of staging all of the required data through the host just to perform these minor computations.
With all computation now on the GPU, ReFacTo can capitalize on direct GPU communication hardware and software, such as those described in Section \ref{sec:gpuComm}.

To utilize CUDA-aware MPI, the only modifications required were to eliminate the device-to-host (DtoH) and host-to-device (HtoD) transfers prior to the communication.
For NCCL, we recreated \texttt{Allgatherv} with a series of calls to \texttt{ncclBcast}, which is the NCCL analogue of \texttt{MPI\_bcast}, since NCCL currently does not have an \texttt{Allgatherv} routine.
Listing \ref{lst:ncclBcast} presents our implementation of \texttt{Allgatherv} for NCCL, which is executed by each GPU.
\texttt{buf} is a buffer stored on each GPU and \texttt{TYPE} refers to the data type of the elements in \texttt{buf} (e.g., ncclChar, ncclInt, etc.).
For a given GPU with rank/ID $i$, if $g$ is equal to $i$, then GPU $i$ will broadcast its portion of \texttt{buf} to all other GPUs.
Otherwise, GPU $i$ will be receiving GPU $g$'s portion of \texttt{buf}. 
We reuse the \texttt{rdispls} and \texttt{recvcounts} arrays from the MPI \texttt{Allgatherv} to determine where each GPU's partition in \texttt{buf} resides and its corresponding length.
Unlike the MPI \texttt{Allgatherv} where there is a send and receive buffer, our NCCL-based implementation uses a single buffer, as per the requirements of \texttt{ncclBcast}, for both sending and receiving data.
This requires that \texttt{buf} for a given GPU be populated with the outgoing data for that GPU, placed in the correct location as dictated by the \texttt{rdispls} array, prior to performing the communication.
Once all the GPUs have synchronized (line 5), \texttt{buf} will hold identical data on all GPUs.
While this implementation incurs some overhead for repeated API calls, it is the most efficient option short of modifying NCCL's source code to implement \texttt{Allgatherv}.
\noindent\begin{minipage}{\linewidth}
\begin{lstlisting}[style=CStyle,label={lst:ncclBcast}, caption={Allgatherv via ncclBcast},columns=flexible]
for (int g = 0; g < numGPUs; g++) {
	ncclBcast(buf+rdispls[g], recvcounts[g], TYPE, 
            g, comm, stream);
}
cudaStreamSynchronize(stream);
\end{lstlisting}
\end{minipage}

We also modified the internals of ReFacTo to support execution on multi-GPU nodes such as the NVIDIA DGX-1 and Cray CS-Storm.
For a cluster with a single GPU per node, each MPI rank is bound to a separate node and utilizes the available GPU on that node, requiring no logic to determine which GPU should be used.
However, on the DGX-1 and CS-Storm, each MPI rank is bound to a separate process and all of the system's GPUs are visible to that process.
Therefore, we associate rank $i$ with the GPU with device ID $i$.
Furthermore, we added the capability to associate the MPI ranks with specific GPUs, allowing for more flexibility on systems where a sequential assignment would not be optimal.


\section{Related Work}
\label{sec:related}
There have been several efforts to evaluate and improve the performance of GPU communication.
Bureddy et al.~\cite{MVAPICH_Benchmarks} extended the popular OSU Micro-Benchmarks (OMB) suite to evaluate the performance of the MVAPICH and OpenMPI CUDA-aware libraries for both point-to-point and collective communication routines.
However, to the best of our knowledge, the irregular collectives are not evaluated with different sized messages per rank in the OMB suite, which does not align with the focus of this work.
Tr{\"a}ff et al.~\cite{irregularAllgather} benchmark an MPI \texttt{Allgatherv} implementation across a set of different message size distributions, but was restricted to host-based communication.
Of particular relevance to our work, Awan et al.~\cite{2017arXivPanda} have compared the performance of the broadcast collective in NCCL and an extended version of MVAPICH-GDR for deep learning workloads.
At the time of their work, the current version of NCCL did not support inter-node communication, so that aspect of the study was omitted.
Furthermore, their study consisted of an evaluation of NCCL and MVAPICH-GDR for regular workloads with respect to message sizes, as well as a focus on deep learning applications, which differs from our work.

In regards to tensor factorization, designing high performance implementations for CP-ALS, as well as measuring their performance, is an active area of research~\cite{rolinger2017JPDC}.
There have been efforts to perform tensor factorization on both shared and distributed memory systems~\cite{smith2015splatt,smith2016medium,Kaya:2015:SST:2807591.2807624}, as well as on GPUs ~\cite{IA32016GPU,GPUCluster17}.
However, to the best of our knowledge, ReFacTo is the only current implementation of CP-ALS that runs on multiple GPUs in a distributed fashion and is able to utilize GPU communication hardware and software.

\section{Performance Evaluation}
\label{sec:eval}
In this section, we describe our experimental setup and present the performance results for three communication libraries on different multi-GPU systems.
We first show the results of the communication libraries on each system for the OSU \texttt{Allgatherv} benchmark, which uses uniform message sizes between the MPI ranks.
These results serve as a baseline of collective communication performance for the different systems and libraries, as well as provide insight into the influence of the different GPU network topologies on communication performance.
We then present the \texttt{Allgatherv} performance of ReFacTo across a suite of data sets with the same library/system combinations as the benchmark experiment.
While the benchmark experiments generate messages of fixed size, the tensor data sets represent real-world applications and have irregular message sizes.

\subsection{Multi-GPU Systems}
\label{sec:setup}
%
%
\begin{figure*}
\centering
\includegraphics[scale=0.30, trim=0cm 0cm 0cm 0cm]{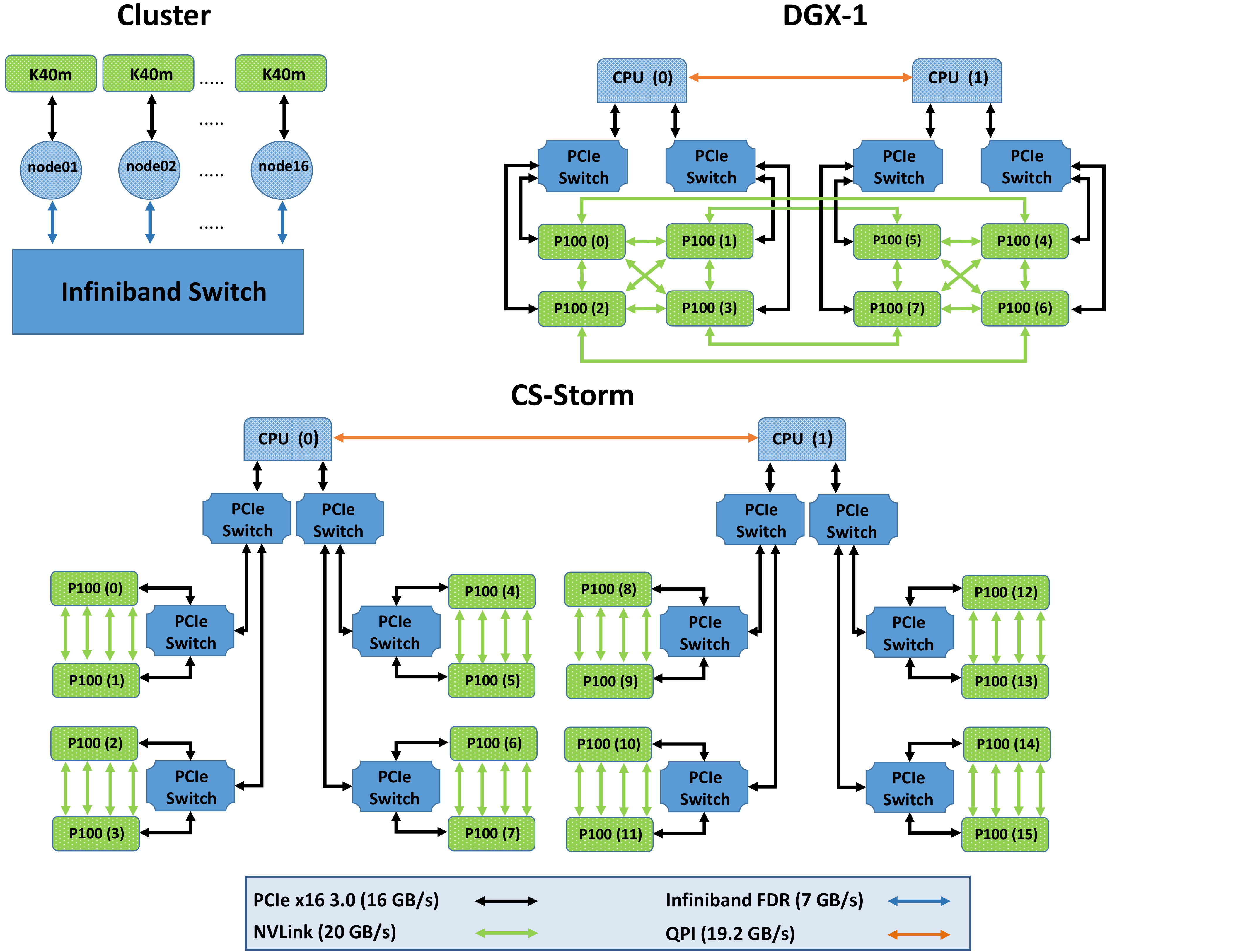}
\caption{Details of the three multi-GPU systems used in our experiments. All reported bandwidths are unidirectional. For the CS-Storm, pairs of GPUs are connected by bonded sets of 4 NVLink connections for a maximum unidirectional bandwidth of 80 GB/sec.}
\label{fig:diagrams}
\end{figure*}

In our experiments, we execute on three different systems: a traditional Infiniband cluster, a NVIDIA DGX-1 and a Cray CS-Storm.
We describe each system below and provide more information in Figure \ref{fig:diagrams}.

The cluster consists of 16 nodes connected by a 56 Gbit/s FDR Infiniband network.
Each node has 512GB of DDR4 DRAM and two 10 core E5-2650v3 Xeon ``Haswell" processors.
The GPU available to each node is a NVIDIA Tesla K40m (Kelper) with 12GB of global memory and 48KB of shared memory per streaming multiprocessor (SM).
Each GPU is connected to its host via PCIe x16 Gen 3.0.
The nodes in the cluster are connected in a star topology.
Figure \ref{fig:diagrams} provides an illustration of this system, denoted as ``Cluster".

NVIDIA's DGX-1 system consists of 8 Tesla P100 (Pascal) GPUs, each with 16GB of global memory and 48KB of shared memory per SM.
The GPUs are connected via NVLink in a hybrid cube-mesh network, where each GPU has four NVLink connection points.
Each NVLink connection point provides point-to-point connection to another GPU at a peak unidirectional bandwidth of 20 GB/sec.
The DGX-1 has 512GB of DDR4 DRAM system memory and two 20 core E5-2698v4 Xeon ``Broadwell" processors.
Figure \ref{fig:diagrams} provides an illustration of this system, denoted as ``DGX-1".

The CS-Storm system used in these experiments consists of 16 Tesla P100 (Pascal) GPUs, each with 32GB of global memory and 48KB of shared memory per SM.
Pairs of GPUs are connected via a bonded set of 4 NVLinks for a peak unidirectional bandwidth of 80 GB/sec.
The pairs of GPUs are also connected to the host via a series of PCIe switches, which facilitate communication amongst the GPUs that are not connected by NVLink.
The CS-Storm has 512GB of DDR4 DRAM system memory and two 18 core E5-2697v4 Xeon ``Broadwell" processors.
An illustration of the CS-Storm is also provided in Figure \ref{fig:diagrams}.

On each system, we evaluate the performance of three different communication libraries: MPI, CUDA-aware MPI and NCCL.
On all systems, the code under test was built using CUDA 8.0.
On the cluster, the CUDA-aware experiments used MVAPICH-GDR 2.2-4 to fully utilize GDR for inter-node communication.
On the other systems, the CUDA-aware experiments used MVAPICH2 2.3b with CUDA support enabled since communication on these systems is restricted to intra-node and GDR is not used.
MVAPICH with CUDA support disabled was used for our traditional MPI experiments.
For NCCL, we used version 2.0.5 on all systems.

\subsection{OSU Benchmark Results}
\label{sec:benchmark}
%
%
%
\begin{figure*}
\centering
\includegraphics[scale=0.58]{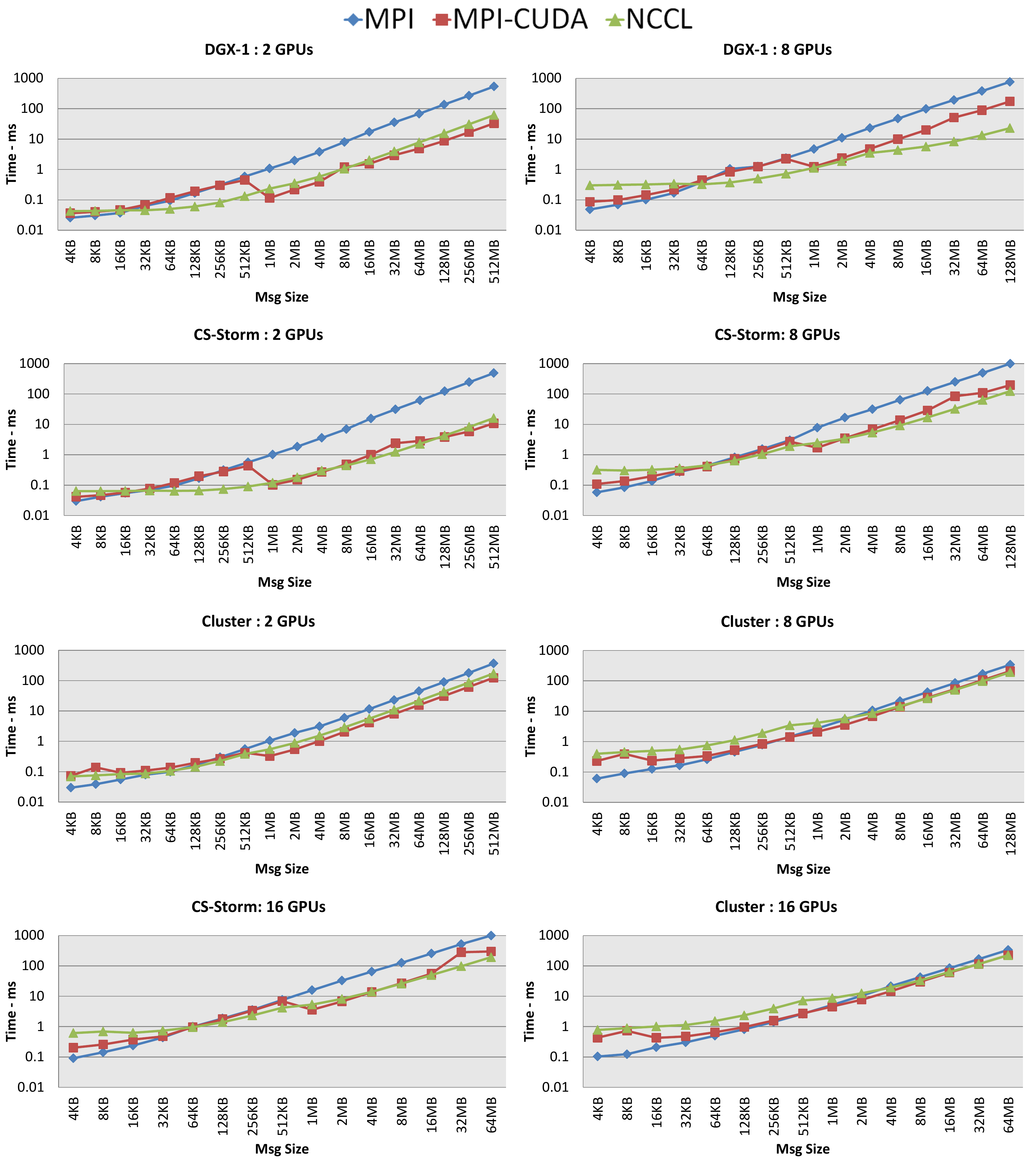}
\caption{Total communication time in milliseconds for the OSU \texttt{Allgatherv} benchmark across the different systems and communication libraries. Note that the vertical axis is logarithmic. The x-axis refers to the size of the messages sent by each rank. The NCCL results are those of a series of broadcasts as NCCL does not support \texttt{Allgatherv}.}
\label{fig:benchmarks}
\end{figure*}

The OSU benchmark for \texttt{Allgatherv} sends fixed size messages to and from every rank.
For a given message size \textit{M} and number of processes \textit{N}, the total communication volume is equal to $M\times N$.
While this does not model irregular workloads, it does provide a baseline for the performance of \texttt{Allgatherv} in the different communication libraries across the systems.
For these experiments, we chose the total maximum communication volume to be 1024MB.
Therefore, we varied the message size from 4KB up to $(1024 / N)$MB, where $N$ is the number of GPUs we used.

As described in Section \ref{sec:refactEXT}, we use a custom implementation of \texttt{Allgatherv} for NCCL as it currently does not support irregular collective communication.
We measure total communication time as the time required to complete the \texttt{Allgatherv} procedure for a given message size, including the time to move the data between the host and GPUs, when applicable.
Therefore, the MPI without CUDA results include the time for the explicit HtoD/DtoH transfers.
We refer to the results for MPI with and without CUDA-support as \textit{MPI-CUDA} and \textit{MPI}, respectively, and the results using NCCL as simply \textit{NCCL}.
%
%
\begin{table*}[t!]
\renewcommand{\arraystretch}{1.0}
\caption{Properties of Data Sets}
\label{tab:datasets}
\centering
\begin{tabular}{|c|c|c|c|c|c|c|c|c|}
\hline
\multicolumn{1}{|c|}{\textbf{Name}} & \multicolumn{1}{c|}{\textbf{Dimensions}}  & \multicolumn{1}{c|}{\textbf{Non-Zeros}} & \multicolumn{2}{c|}{\textbf{Avg Msg Size}} & \multicolumn{2}{c|}{\textbf{Min/Max Msg Size}} & \multicolumn{2}{c|}{\textbf{Msg Size CV}}\\
\cline{4-9}
 & & & 2 GPUs & 8 GPUs & 2 GPUs & 8 GPUs & 2 GPUs & 8 GPUs\\
\hline
NETFLIX    & 480K x 18K x 2K   & 100M  & 6.4MB   & 1.6MB  & 0.04MB / 26.5MB  & 0.01MB / 13.5MB   & 1.5  & 1.84\\
AMAZON     & 524K x 2M x 2M    & 200M  & 65.2MB  & 16.3MB & 24.6MB / 89.5MB  & 5.9MB / 23.7MB    & 0.44 & 0.44\\
DELICIOUS  & 532K x 17M x 2M   & 140M  & 128.9MB & 32.2MB & 0.2MB / 496.2MB  & 0.006MB / 152.4MB & 1.35 & 1.48\\
NELL-1     & 3M x 2M x 25M     & 143M  & 291.3MB & 72.8MB & 61.3MB / 729.8MB & 14.7MB / 183.5MB  & 1.06 & 1.06\\
\hline
\end{tabular}
\end{table*}
Figure \ref{fig:benchmarks} presents the results of the OSU benchmark across the three systems when using the different libraries and up to 16 GPUs when applicable.
The x-axis denotes the per-rank message size and the y-axis is the total communication time in milliseconds.
Note that the y-axis is logarithmic.
On the DGX-1 and CS-Storm for messages larger than 16KB, both NCCL and MPI-CUDA outperform traditional MPI by a significant margin when using 2 GPUs due to leveraging GPUDirect P2P via NVLink between the GPUs.
The difference is much greater on the CS-Storm since there is a bonded set of 4 NVLink connections between the two GPUs.
On the cluster, NCCL and MPI-CUDA still outperform MPI when using 2 GPUs but by a much smaller factor and only after the message sizes are 256KB or larger.
Since the cluster has only Infiniband between nodes, NCCL and MPI-CUDA are limited by similar constraints as traditional MPI and receive at most a 2.5x improvement over MPI due to more efficient copying between the hosts and GPUs as well as from using GDR.

When using 8 GPUs on the DGX-1, NCCL provides faster runtimes over MPI-CUDA for messages larger than 64KB due to the unique GPU network topology of the DGX-1.
On the DGX-1, any GPU can be reached by another with at most two NVLink hops.
Since NCCL does not rely entirely on the availability of P2P access between GPUs, it can take advantage of the DGX-1's topology and use NVLink exclusively for communication between all 8 GPUs.
On the other hand, MVAPICH with CUDA-support cannot infer that some GPUs can be reached by traversing two NVLink connections and defaults to the PCIe topology for that portion of the collective communication.
On the CS-Storm, we observe that NCCL also provides better performance over MPI-CUDA but only when the message sizes are larger than 4MB.
However, the difference in performance is not as significant as on the DGX-1.
This is because only pairs of GPUs in the CS-Storm are connected via NVLink, so NCCL does not have the same advantage as it does on the DGX-1.
On the cluster when using 8 GPUs, we can see that all three libraries begin to exhibit similar performance as the message sizes become larger, with NCCL and MPI-CUDA still performing slightly better than traditional MPI.

It is worth mentioning the sudden decrease in runtime for MPI-CUDA across the systems once the message sizes reach 1MB.
This suggests that there are one or more parameters within MVAPICH whose values can have noticeable influence on its performance.
The MVAPICH documentation notes that the optimal values for such parameters are selected based on architecture detection.
However, determining the optimal values for these parameters can become a difficult task when the sizes of the messages are highly irregular, as we will demonstrate in Section \ref{sec:results}.

In terms of the systems themselves, the CS-Storm offers the best performance for both MPI-CUDA and NCCL when using 2 GPUs while the DGX-1 provides the fastest runtimes for NCCL and MPI-CUDA when using 8 GPUs.
We observe that NCCL on the DGX-1 is up to 8.3x faster than NCCL on the cluster, highlighting the benefit of a specialized dense multi-GPU system like the DGX-1.
When leveraging 16 GPUs on the CS-Storm and cluster, we observe trends similar to the results on 8 GPUs.
However, the runtime of the MPI libraries on the cluster when using 16 GPUs are as much as 4.5x faster than the CS-Storm.
While pairs of GPUs on the CS-Storm are connected by NVLink, all other communication must utilize the PCIe fabric.
With GPUs sharing PCIe switches, this results in poor performance when compared to the cluster, where each GPU has exclusive access to its local PCIe bus.
%
%
%
\begin{figure*}
\centering
\includegraphics[scale=0.58]{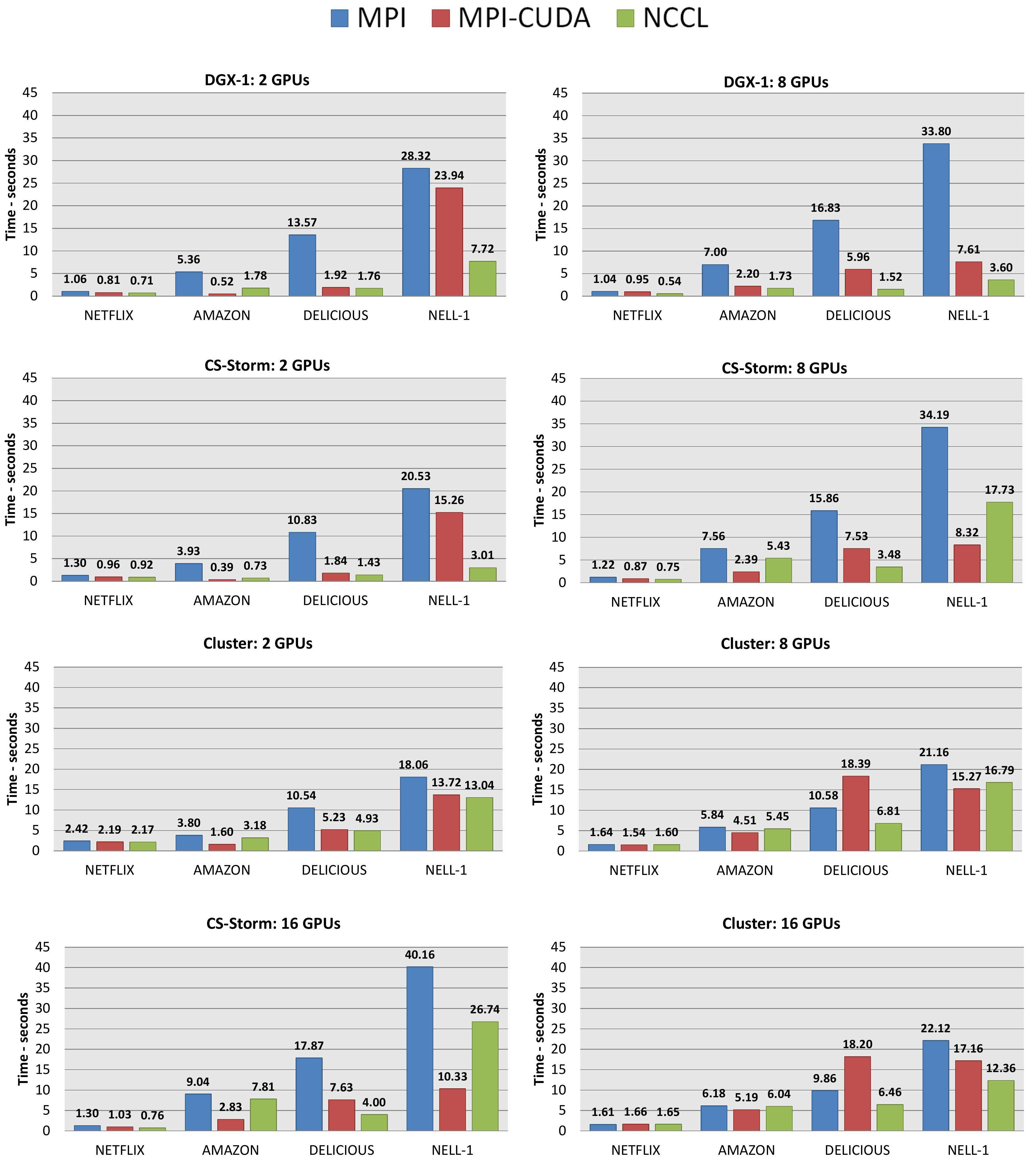}
\caption{Total communication time in seconds for ReFacTo across the data sets on the different systems and communication libraries. Shorter bars represent better performance.}
\label{fig:tensorRes}
\end{figure*}
\subsection{Tensor Factorization Results}
\label{sec:results}
While the results from the OSU benchmark provide insight into the performance of \texttt{Allgatherv} for the three systems and communication libraries, the message sizes are regular.
As the goal of this work is to investigate how communication performance is impacted by irregular workloads, we conducted an empirical evaluation of the communication performance in ReFacTo, the distributed and multi-GPU sparse tensor factorization tool described in Section \ref{sec:tensors} that uses \texttt{Allgatherv}.

For the following results, ReFacTo was built using CUDA 8.0, NCCL 2.0.5, 32-bit integers and single-precision floating point numbers on all three systems.
Similar to the OSU benchmark experiment, the communication runtime of ReFacTo is measured on the three systems while using the different communication libraries.
We focus our measurements on the time required to perform all of the GPU communication during the tensor factorization, including HtoD/DtoH transfers when applicable.
As in the previous section, we refer to the results for MPI with and without CUDA-support as \textit{MPI-CUDA} and \textit{MPI}, respectively, and the results using NCCL as simply \textit{NCCL}.

Table \ref{tab:datasets} presents the data sets that we evaluate in our experiments for ReFacTo and are shown in ascending order with respect to the average message size, assuming single-precision floating point numbers.
These data sets represent real-world applications and are not synthetically generated.
While the communication is collective, we consider the size of all the messages sent by reach rank throughout the factorization of the data sets.
It is worth noting the significant difference between the smallest and largest message sizes for each data set, which can be as large as 25,400x in the case of DELICIOUS.
Furthermore, the difference in message sizes within a single \texttt{Allgatherv} call for DELICIOUS is as large as 13,500x.
We also show the coefficient of variation (CV) of the message sizes for each data set, which is the ratio of the standard deviation to the mean and is a measure of relative variability.
As it can be seen, the CV for these data sets are significant and illustrate the high degree of irregularity in the message sizes throughout the factorization. 
The AMAZON, DELICIOUS, and NELL-1 data sets can be obtained from the Formidable Repository of Open Sparse Tensors and Tools (FROSTT)~\cite{frosttdataset}.
The AMAZON data set was modified in our experiments to only include 200 million of its 1.7 billion non-zeros.
The NETFLIX data set contains movie review data from the Netflix Prize Competition~\cite{Bennett07thenetflix}.
When using MVAPICH-GDR, we performed a parameter sweep for the \texttt{MV2\_GPUDIRECT\_LIMIT} variable on each data set to determine the best value.
This value determines when GDR is used by MVAPICH and is recommended by the developers to be tuned based on the system's node architecture, processor, GPU and Infiniband card.

Figure \ref{fig:tensorRes} presents the total communication runtime of ReFacTo across the data sets on the different systems and when using the different communication libraries.
The y-axis represents total communication time in seconds.
The data sets are labeled on the x-axis and are ordered from smallest to largest with respect to average message size.

We observe that NCCL on the DGX-1 is up to 4.7x faster than NCCL on the cluster.
For an application that depends on collective communication between GPUs, the DGX-1 provides an ideal topology where every GPU can be reached by any other GPU in at most two NVLink hops.
A GPU network topology like that of the cluster does not lend itself well to such applications, as there are no direct paths between the GPUs other than through the compute nodes and Infiniband network.
However, there are instances where the cluster is better suited than the specialized CS-Storm system, much like we observed in the benchmark results.
We see as much as a 2.16x difference in performance for NCCL on the cluster and CS-Storm when using 16 GPUs and as much as a 1.81x difference between both MPI and MPI-CUDA.

NCCL was designed for deep learning applications, where the message sizes are typically regular within the collective communication.
Furthermore, NCCL's targeted architectures are specialized dense multi-GPU systems.
However, the results in Figure \ref{fig:tensorRes} offer evidence that NCCL is also competitive with more general-purpose communication libraries on traditional systems.
We observe that NCCL on the cluster is 1.2x faster on average than MPI-CUDA across the tensor data sets and number of GPUs.
Also, the communication workloads in tensor factorization are vastly different than those of deep learning applications, further exhibiting the viability of NCCL as a general purpose GPU communication library.

Contrary to the benchmark results in Figure \ref{fig:benchmarks}, NCCL on all of the systems when using two GPUs exhibits better performance than MPI-CUDA across all of the tensors with the exception of AMAZON.
In the case of the NELL-1 data set, whose message sizes range from 61.3MB to 729.8MB, NCCL is 3.1x faster than MPI-CUDA on the DGX-1.
However, on the benchmark for similar message sizes, NCCL is 1.76x slower on average than MPI-CUDA on the DGX-1.
On the CS-Storm, NCCL is 5x faster than MPI-CUDA when utilizing two GPUs for the NELL-1 data set, yet the benchmark results for the CS-Storm when using two GPUs show that NCCL is outperformed by MPI-CUDA by as much as 1.5x for large message sizes.
On the cluster when using two GPUs, we observe similar trends as those described above but with much smaller gaps in performance.
This is because on the cluster, all of the communication libraries use the same hardware for communication, namely PCIe and Infiniband.
On the other hand, the DGX-1 and CS-Storm have more complicated topologies that involve NVLink, PCIe, and QPI, resulting in much larger performance gaps due to how each library utilizes the topology and hardware.

From the benchmark results, we can see that MPI-CUDA's performance is strictly worse on 8 GPUs compared to two GPUs.
This is also generally true for the results in Figure \ref{fig:tensorRes}.
However, the performance of MPI-CUDA on the NELL-1 data set when using 8 GPUs on the DGX-1 improves by 3.14x when compared to the results on two GPUs.
On the CS-Storm when using 8 GPUs for the NELL-1 data set, MPI-CUDA's communication runtime is 1.83x faster than when using two GPUs.
Such performance improvements were not seen in the benchmark results and underscore the difficulties of predicting performance that involves irregular behavior.

While the results from the benchmark suggest that MPI-CUDA is strictly faster than traditional MPI on the cluster for message sizes larger than 256KB, the DELICIOUS data set contradicts this trend.
When using two GPUs, MPI-CUDA is faster than traditional MPI, which would be expected based on the benchmark results and the range of message sizes for the DELICIOUS data set.
However, when the number of GPUs is increased to 8, we observe that MPI-CUDA is now 1.73x slower than traditional MPI and when using 16 GPUs, it is 1.85x slower.
Among all of the data sets in Table \ref{tab:datasets}, the DELICIOUS data set consists of the largest range of message sizes.
We found that because of this high irregularity, the communication runtime for the DELICIOUS data set was sensitive with respect to the \texttt{MV2\_GPUDIRECT\_LIMIT} variable.
For example, we observed a 3.1x difference in communication runtime between a 1MB and 4MB value for this variable.
Furthermore, the optimal value for this variable can vary significantly when using a different number of GPUs.
While 512MB is the optimal value for the DELICIOUS data set when using two GPUs, we found that 16B is the optimal value when using 8 GPUs, which is $2^{25}$ times smaller.
We suspect that because of the wide range of message sizes within the DELICIOUS data set, MPI-CUDA suffers on the cluster due to buffer size limitations for GDR.
Note that we do not see this behavior on the other systems, which do not use GDR.

\section{Conclusion}
\label{sec:concl}

Our findings are summarized below and align with the questions posed in Section \ref{sec:intro}:

\begin{itemize}
\item We found as much as a 8.3x difference in \texttt{Allgatherv} runtime between the DGX-1 and cluster when using NCCL on the OSU benchmark. On the tensor data sets, we observed as much as a 4.7x difference. These results highlight the benefits of a specialized multi-GPU system over a more traditional cluster.
\item NCCL was found to be 1.2x faster on average than MVAPICH-GDR on the cluster for the tensor factorization experiment, despite being designed for dense multi-GPU systems and deep learning applications with regular workloads. This highlights the viability of NCCL as a general-purpose GPU communication library.
\item We found that irregular message sizes resulted in performance trends in the tensor factorization results that were not shown in the OSU benchmark. Furthermore, the performance of MVAPICH-GDR was shown to be highly sensitive with respect to the values of the \texttt{MV2\_GPUDIRECT\_LIMIT} variable for very irregular data sets. These findings underscore the complexities of irregular workloads in collective GPU communication.
\end{itemize}

For future work, we would like to implement an \texttt{Allgatherv} routine within NCCL, or evaluate the routine if it is released in a new version, instead of relying on a series of broadcasts calls.
We are also interested in evaluating communication performance for other irregular applications and on systems with more GPUs per node.
Furthermore, it would be worthwhile to incorporate the message size distribution benchmarks developed by Tr{\"a}ff et al.~\cite{nonUniformMPI} into a GPU-based benchmark.

\bibliographystyle{IEEEtran}
\bibliography{CCGRID_Refs} 

\end{document}